\documentclass[iop]{emulateapj}
\usepackage{graphicx,apjfonts}

\begin{document}

\title{A possible signature of Lense-Thirring precession in dipping and eclipsing neutron-star low-mass X-ray binaries}

\author{Jeroen Homan}
\affil{Kavli Institute for Astrophysics and Space Research, Massachusetts Institute of Technology \\ 70 Vassar Street, Cambridge, MA 02139, USA}
\email{jeroen@space.mit.edu}

\shorttitle{Signatures of Lense-Thirring precession}
\shortauthors{Homan}

\begin{abstract}

Relativistic Lense-Thirring precession of a tilted inner accretion disk around a compact object has been proposed as a mechanism for low-frequency ($\sim$0.01--70 Hz) quasi-periodic oscillations (QPOs) in the light curves of X-ray binaries. A substantial misalignment angle ($\sim$15--20$^\circ$) between the inner-disk rotation axis and the compact-object spin axis is required for the effects of this precession to produce observable modulations in the X-ray light curve. A consequence of this misalignment is that in high-inclination X-ray binaries the precessing inner disk will quasi-periodically intercept our line of sight to the compact object. In the case of neutron-star systems this should have a significant observational effect, since a large fraction of the accretion energy is released on or near the neutron-star surface. In this Letter I suggest that this specific effect of Lense-Thirring precession may already have been observed as $\sim$1 Hz QPOs in several dipping/eclipsing neutron-star X-ray binaries. 

\end{abstract}

\keywords{accretion, accretion disks --- relativistic processes --- X-rays: binaries}

\section{Low-frequency QPOs}\label{sec:intro}

Low-frequency quasi-periodic oscillations (LF-QPOs) have been observed in the X-ray light curves of most low-mass X-ray binaries \citep{va2006}. Their frequencies are typically $\sim$1--70 Hz for neutron-star low-mass X-ray binaries (NS-LMXBs) and $\sim$0.01--30 Hz for black-hole low-mass X-ray binaries (BH-LMXBs). Despite having been known for more than 25 years, their origin is still not well understood. However, similarities between the LF-QPOs in NS- and BH-LMXBs suggest that they have a common origin in the inner accretion flow around the compact object \citep{wiva1999,psbeva1999}; these similarities are most obvious between the so-called type-C QPOs in BH-LMXBs \citep{resomu2002}, the horizontal branch oscillations in Z-type NS-LMXBs and the ``low-frequency QPO'' in atoll-type LMXBs \citep{va2006}. Hereafter, ``LF-QPO''  will be referring to these specific types of LF-QPOs in LMXBs. These LF-QPOs  are all accompanied by a peaked noise component at frequencies that are typically less than one decade lower than that of the QPO. Both the LF-QPOs and the accompanying noise are typically seen in intermediate spectral states. 

Among the most promising models for LF-QPOs are the ones that invoke Lense-Thirring precession. This is a relativistic effect that occurs as the result of the dragging of inertial frames around a spinning compact object, when the rotation axis of a particle orbit is inclined with respect to the spin axis of the compact object \citep{leth1918,bape1975}. \citet{stvi1998} were the first to suggest Lense-Thirring precession as an explanation for the LF-QPOs in NS-LMXBs;  \citet{stvimo1999} also applied the model to LF-QPOs in BH-LMXBs. 

Initial work on these models focussed on reproducing the observed QPO frequencies and only considered motion of particles or ``blobs'' along infinitesimally perturbed orbits, but no clear mechanism was proposed that could explain the actual modulations in the X-ray flux. Simulations by \citet{schomi2006} of a tilted precessing ring of gas around a spinning black hole show that such modulations can naturally arise from relativistic Doppler and light bending effects, without relying on the existence of inhomogeneities in the accretion flow. However, substantial misalignment angles (15-20$^\circ$) between the compact object's spin axis and the rotational axis of the ring are needed to reproduce the observed QPO amplitudes. Recent work by \citet{indofr2009} and \citet{indo2011a,indo2011b} describes the Lense-Thirring precession of a geometrically thick inner accretion flow. These authors also require a substantial misalignment ($\sim$15$^\circ$) in order to get a density profile that results in QPO frequencies that match the observed ones; without such a misalignment, the predicted frequencies are too high and the dependence on the spin of the compact object is too strong.

A problem to these precession models is presented by recently discovered 35--50 Hz QPOs in an accreting 11 Hz pulsar \citep{alinva2012}. These frequencies are too high (by factors of at least 40) for Lense-Thirring precession in such a slowly spinning source. The strong similarity of these QPOs to the horizontal branch oscillations in the Z sources casts doubt on the Lense-Thirring interpretation of that type of LF-QPOs and, by extension, the LF-QPOs in atoll sources and BH-LMXBs as well.

The large misalignment angles that are required in the precession models described above may be problematic as well: it is not clear if such angles can be maintained for a long enough period. \citet{bape1975} showed that a misaligned thin disc should gradually align itself with a black hole at small radii. \citet{frblan2007} performed simulations of a thick flow  around a spinning black hole (similar to that used by \citealt{indofr2009}), incorporating the effects of the black hole space-time as well as magnetorotational turbulence. They find no indications for the Bardeen-Petterson effect, suggesting that misaligned inner accretions flows may persists for long times. In this Letter I present observational evidence in support of misaligned inner accretion flows in dipping/eclipsing NS-LMXBs.

\begin{figure}[t]
\epsscale{0.9}
\plotone{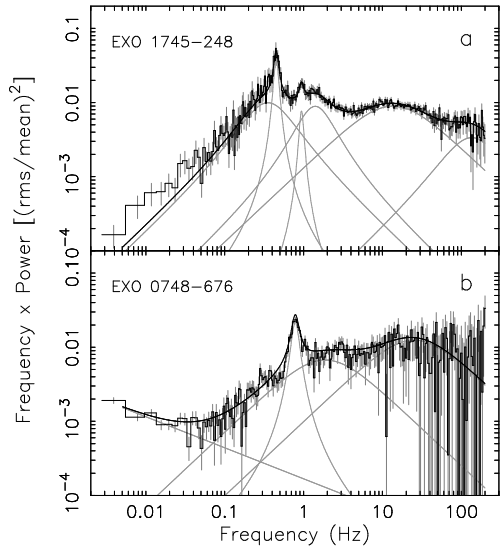}
\caption{Examples of $\sim$1 Hz QPOs. Panel (a) shows a power spectrum of EXO 1748--248 with a QPO at 0.44 Hz and panel (b) shows a power spectrum of EXO 0748-676 with a QPO at 0.78 Hz. The best fits (black curves) to the power spectra are shown, together with the individual components (gray curves). In both cases the QPO is accompanied by significant peaked noise components (with at least one being close to the QPO frequency), while in EXO 0748--676 power-law noise was present as well. The {\it RXTE} observation used for panel (a) was 50054-06-04-03; for panel (b) observations 20082-01-[01 \& 02] were used. The power spectra were created and fitted following procedures outlined in \citet{homiwi2005}.} 
\label{fig:pds}
\end{figure}

\section{QPOs in dipping/eclipsing NS-LMXBs}

The large misalignment angles required to reproduce the observed LF-QPO amplitudes and frequencies have a potentially observable consequence that has not been discussed in previous works: if the inclination angle of a source is within a certain range, part of the titled and precessing (thick) inner accretion flow could quasi-periodically intercept the line of sight between an observer and the central compact object. For a misalignment angle $\alpha$ this is expected to happen in sources that are viewed at inclination angles that are larger than $90^\circ-\alpha$; for $\alpha\approx20^\circ$ \citep{schomi2006,indofr2009} this means inclination angles larger than $70^\circ$. Given the absence of a solid surface in BH-LMXBs, the movement of the precessing  disk through our line of sight should not result in substantial flux modulation of the compact object itself in those sources, although, as pointed out by \citet{indofr2009}, self-occultation of the disk may occur. However, for NS-LXMBs, where up to half of the accretion energy is released on the neutron star surface or in the boundary layer, substantial modulations will arise when the precessing disk moves through our line of sight. 

As I discuss below, evidence for such modulations has been seen in a number of dipping and/or eclipsing NS-LMXBs. The dips and eclipses in the X-ray light curves of these sources are the result of absorption by extended features in the outer accretion disk and obscuration by the companion star, respectively. Dips are typically seen in sources with inclination angles $>60^\circ$, whereas full (i.e.\ not partial) eclipses are seen in sources with inclination angles between 75$^\circ$ and 80$^\circ$ \citep{frkila1987}. These sources are therefore expected to show the effects of the inner accretion disk moving through our line of sight to the compact object if the misalignment angle $\alpha$ is indeed on the order of $15^\circ-20^\circ$.

The first dipping NS-LMXB in which clear evidence was found for modulations of the neutron-star emission is XB 1323--619. \citet{jovawi1999} discovered a 0.77--0.87 Hz QPO in data obtained with the {\it Rossi X-ray Timing Explorer (RXTE)}, with an amplitude that was found to be consistent with being constant during the persistent emission, X-ray dips, and, most importantly, type I X-ray bursts (which are the result of thermonuclear burning on the neutron-star surface). Unlike LF-QPOs in non-dipping/eclipsing NS-LXMBs, the fractional rms amplitude of this QPO depended only weakly on energy. Based on these properties \citet{jovawi1999} concluded that the QPO was caused by quasi-periodic obscuration of the central X-ray source by a structure in or on the accretion disk. The fact that the $\sim$1 Hz QPOs persist throughout dips with nearly constant rms amplitudes indicates that they are formed at a radius smaller than the radius at which the dipping structures are located.

Soon after, a QPO with similar properties was found in the dipping/eclipsing NS-LMXB EXO 0748--676 by \citet{hojowi1999}. The frequency of this QPO was found to be more variable (0.58--2.44 Hz), but otherwise its properties were remarkably similar. Despite the observed frequency range, I will refer to this type of QPO as the ``$\sim$1 Hz QPO'' throughout the remainder of this Letter, to set it apart from the LF-QPOs in the non-dipping/eclipsing NS-LMXBs. An example of a $\sim$1 Hz QPO from EXO 0748-676 is shown in Figure \ref{fig:pds}b. The only observation in which \citet{hojowi1999} did not find the $\sim$1 Hz QPO had a count rate 2--3 times higher than the other observations in their data set. Analysis of a larger data set by \citet{hova2000} increased the frequency range to 0.4--3 Hz and also made it clear that the observation without the $\sim$1 Hz QPO corresponded to the spectrally soft state, whereas the observations with  $\sim$1 Hz QPOs were all in spectrally harder states. A preliminary inspection of more recent EXO 0748--676 data reveals $\sim$1 Hz QPOs up to at least 15 Hz.

\begin{table}[t]
\begin{center}
\caption{Dipping/Eclipsing NS-LMXBs Observed with {\it RXTE}}\label{tab:sources}
\vspace{0.1cm}
\begin{tabular}{lccc}
\hline
\hline
Source & $\sim$1 Hz QPOs & Refs\\
\hline
EXO 0748--676      		& Yes 	&  1,2 	\\
XB 1254--690$^\star$		& No 			& 3		\\
XTE J1710--281     		& Yes 	& 4		\\
4U 1746--37          	& Yes 	& 5		\\ 
MXB 1659--298       		& ? 			& 		\\
4U 1624--49$^\star$		& No 			& 6 		\\
XB 1323--619        		& Yes 	& 7 		\\
4U 1915--05         		& ? 			& 8		\\
XTE J1759--220     		& Yes 	& 9 		\\
1A 1744--361       		& Yes 	& 10 		\\
EXO 1745--248    		& Yes 	& 11, 12		\\ 
\hline
\end{tabular}
\end{center}
\noindent $^\star$ These sources have only been observed in the soft state\\

References: {\bf 1} \citet{hojowi1999} {\bf 2} \citet{hova2000} {\bf 3} \citet{bh2007}  {\bf 4} Preliminary analysis {\bf 5} \citet{jovaho2000} {\bf 6} \citet{lovava2005} {\bf 7} \citet{jovawi1999} {\bf 8} \citet{bobaol2000} {\bf 9} R.\ Wijnands 2008, private communication {\bf 10} \citet{bhstsw2006} {\bf} {\bf 11} Altamirano et al., in preparation {\bf 12} \citet{mubh2011}

\end{table}

The third source in which a $\sim$1 Hz QPO was found is the dipping NS-LMXB 4U 1746--37. \citet{jovaho2000} detected a 1.01--1.59 Hz QPO that, like the QPOs in XB 1323--619  and EXO 0748--676, persisted through type I X-ray bursts. Also, like in EXO 0748-676, a clear dependence on spectral state was observed; the QPO was only found in the spectrally hard state and was not detected in the spectrally soft state.

\begin{figure}[t]
\epsscale{0.9}
\plotone{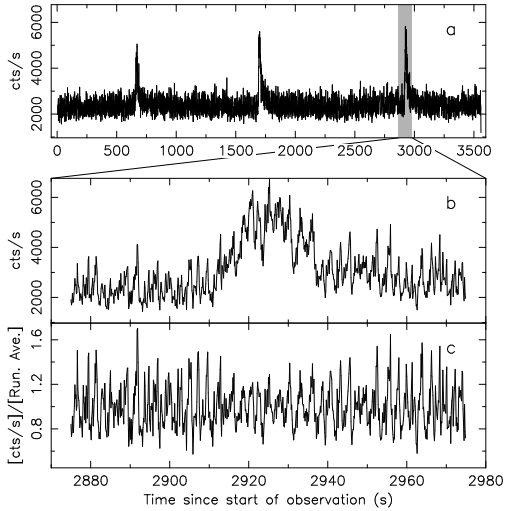}
\caption{(a) Light curve of EXO 1745--248 (observation 50054-06-04-03), showing three type I X-ray bursts. Panel (b) shows a enlargement of the dark shaded area in panel (a). A strong $\sim$0.4 Hz QPO can be seen in the light curve that persists during the type I X-ray burst. Panel (c) shows the light curve from panel (b), divided by a 3.125 s running average. Time resolution is 1s in all three panels. Fits to the power spectra from outside and inside the bursts give fractional rms amplitudes of 11.1\%$\pm$0.6\% and 13.7\%$\pm$2.1\%, respectively.  } 
\label{fig:burst}
\end{figure}

In addition to the three sources discussed above, there are eight other dipping and/or eclipsing NS-LMXBs that have been observed extensively with {\it RXTE} (with total exposure times ranging from $\sim$135 ks to $\sim$660 ks).  These sources are listed in Table \ref{tab:sources}. $\sim$1 Hz QPOs have been found in three of these sources: in 1A 1744--361 \citep{bhstsw2006}, where it was only seen in the hard state, EXO 1745--248 (Altamirano et al., in preparation, \citealt{mubh2011}), also  only in the hard state, and XTE J1759--220 (Rudy Wijnands, private communication; spectral state unknown).  An example power spectrum from EXO 1745--248, showing a $\sim$1 Hz QPO,  can be seen in Figure \ref{fig:pds}a. The corresponding light curve of this observation is shown in Figure \ref{fig:burst};  the $\sim$0.4 Hz QPO can clearly be seen in the light curve, both in the persistent and in the burst emission. 

Hints of $\sim$1 Hz QPOs were also seen in a preliminary analysis of data of XTE J1710--281. In 4U 1624--49 \citep{lovava2005} and XB 1254--690 \citep{bh2007} no $\sim1$ Hz QPOs (or LF-QPOs) were observed; both sources were only found in the soft state. \citet{bobaol2000} report on QPOs between 5 and 80 Hz in  the dipping NS-LMXB 4U 1915-05, but it is not clear if these QPOs have similarities to the $\sim$1 Hz QPO in the other dipping/eclipsing sources. I was not able to find rapid variability studies of the eclipsing NS-LMXB MXB 1659--298. A summary of  $\sim$1 Hz QPO detections is given in Table \ref{tab:sources}.

\section{Discussion}

The $\sim$1 Hz QPOs detected in dipping/eclipsing NS-LMXBs satisfy one of the implicit predictions of the latest versions of the Lense-Thirring precession models for LF-QPOs; they are caused by matter moving through our line of sight to the neutron star. This latter property is deduced from the fact that the strength of the $\sim$1 Hz QPOs remains more-or-less constant during type I X-ray bursts, as can  be seen in Figure \ref{fig:burst}, for example. The fact that the $\sim$1 Hz QPOs are only detected in dipping/eclipsing sources (i.e. sources that are seen at viewing angles larger than $\sim$60$^\circ$) falls in line with the misalignment angles of $\sim$15--20$^\circ$ that are implied by the work of \citet{schomi2006} and \citet{indofr2009}. 

The frequency of the $\sim$1 Hz QPOs was originally interpreted as being Keplerian by \citet{jovawi1999} and \citet{hojowi1999}. For the $\sim$0.4--15 Hz range seen in EXO 0748--676 this would correspond to radii of  $\sim2.8\times10^7$--$3.1\times10^8$ cm (assuming a 1.4 $M_\odot$ NS). However, it is not clear what type of obscuring structures might exist at such radii that could cause the $\sim$1 Hz QPOs. In the Lense-Thirring precession interpretation, the radius at which the modulation of the neutron-star emission occurs is much smaller. Using the Lense-Thirring precession frequency equations from \citet{stvi1998}, and assuming a neutron-star spin of 552 Hz \citep{galich2010}, a mass of 1.4 $M_\odot$, and a moment of inertia of $10^{45}$ g\,cm$^2$, we find typical radii of $\sim1.8\times10^6$--$5.9\times10^6$ cm for EXO 0748--676. The corresponding range for Keplerian frequencies at those radii is $\sim$150--930 Hz, which largely overlaps with the range of upper kHz QPO frequencies (often interpreted as Keplerian) reported for NS-LMXBs \citep{va2006}.

An important property of the $\sim$1 Hz QPOs is their very weak energy dependence. This sets them apart from the LF-QPOs seen in NS-LMXBs and BH-LMXBs, which generally  increase in strength with energy. This weak energy dependence could easily be produced by (partial) obscuration of the central X-ray source by an opaque medium, such as an optically thick accretion disk. However, the part of the flow that is intercepting our line of sight to the central X-ray source in the geometrical model of \citet{indofr2009} is assumed to be a translucent ($\tau\sim1$) Comptonizing medium, rather than an optically thick medium. Still, most of the photons from the central X-ray source that will interact with the Comptonizing medium will be scattered out of our line of sight, and, since the chance of scattering (set by the Thomson scattering cross section) is independent of photon energy, the resulting modulation will not have a strong energy dependence.

In the framework of the Lense-Thirring precession models, both the LF-QPOs and the $\sim$1 Hz QPOs in NS-LMXBs could be the result of the same precession mechanism. For sources with inclinations less than $\sim$60$^\circ$ modulations in the X-ray flux would be the result of relativistic beaming and light bending effects \citep{schomi2006}, whereas for sources with higher inclinations the modulations would be the result of obscuration of the central X-ray source. So, while the frequency of the two types of QPOs is set by the (same) precession frequency, the actual modulation mechanism would depend on the viewing angle. It is therefore interesting to see how the properties of the $\sim$1 Hz QPOs compare with those of the LF-QPOs, also in light of the recent findings by \citet{alinva2012}.

$\sim$1 Hz QPOs are predominantly detected in the spectrally hard state, with some detections in the intermediate state. LF-QPOs are, however, predominantly detected in intermediate spectra states, with occasional detections in hard and soft spectral states.

LF-QPOs increase in frequency as the X-ray spectrum softens. For $\sim$1 Hz QPOs this dependence has not yet been studied, but there are strong indications that the highest frequency $\sim$1 Hz QPOs in EXO 0748--676 are found in the spectrally softer states than the lowest frequency ones.

In terms of QPO frequency range there is a clear discrepancy between the two types of QPOs. Based on their X-ray luminosity and spectra, the dipping/eclipsing NS-LMXBs are almost all atoll-type sources that spend most of their time in the spectrally hard and intermediate states. The frequency ranges for LF-QPOs in atoll sources are typically $\sim$5--50 Hz \citep{vavadi2002,vavame2003,alvame2008}, whereas $\sim$1 Hz QPOs have been found in the $\sim$0.4--15 Hz range. It is not clear what could cause such discrepancy, if both types of QPOs were to be the result of the same precession mechanism. Perhaps the modulation mechanism of the $\sim$1 Hz QPOs is only effective at the lowest precession frequencies, when the precessing part of the inner accretion flow is assumed to have its largest extent. This is supposed to occur in the hardest spectral states, where typically no clear indications of narrow LF-QPOs are seen \citep{vavadi2002,vavame2003,alvame2008}. 

Although previous works on the $\sim$1 Hz QPOs in the dipping/eclipsing sources only reported power-law noise components in addition to the QPO,\footnote{This may have partly been the result of short time segments (16 s) that were used to create the power spectra in these works; for the power spectra in Figure \ref{fig:pds} 256 s segments were used. Also, in the case of EXO 0748--676, which has very low count rates ($\sim$24 counts\,s$^{-1}$\,PCU$^{-1}$), a relatively large amount of data ($\sim$40 ks) had to be combined for the peaked noise to show up significantly.} they are in fact  accompanied by peaked noise as well, as can been from Figure \ref{fig:pds}. This is a characteristic they share with LF-QPOs. However, plotting the frequencies of the QPOs and the (lowest frequency) noise components from Figure \ref{fig:pds} against each other, we find that the $\sim$1 Hz QPOs fall well below the relation traced out by the LF-QPOs in the work of \citet{wiva1999}, by factors of 5--10 in QPO frequency.

Clearly, there are substantial differences between the two types of QPOs. It is not clear how these might be related to differences in viewing angle. Observing $\sim$1 Hz QPOs and LF-QPOs simultaneously (at different frequencies) would indicate that the two types of QPOs are {\em not} the result of the same precession mechanism, since the inner accretion flow is assumed to be precessing as a solid body, with a single frequency \citep{frblan2007,indofr2009}. However, published works on the $\sim$1 Hz QPOs indicate that, except for occasional higher harmonics (see Figure \ref{fig:pds}a), the $\sim$1 Hz QPOs are not accompanied by (other) LF-QPOs.


As discussed in Section \ref{sec:intro}, the Lense-Thirring precession models of \citet{schomi2006} and \citet{indofr2009} require large misalignment angles between the inner accretion disk and the neutron star spin axis. In the presence of accretion, a low-magnetic field neutron star should have its spin axis aligned with that of the binary orbit, leaving no obvious misalignment angle between inner disk and neutron star. There are several effects that may result in such a misalignment. \citet{pr1996} showed that the strong radiation field from the central X-ray source can induce warps in the accretion disk. However, this happens only at relatively large radii ($>3\times10^8$ cm). A more promising source of  misalignment could be magnetically driven disc warping, in which the weak ($\sim10^8$ G) tilted magnetic field of the neutron star induces a disk tilt \citep{la1999}. In fact, \citet{shla2002} provide a detailed discussion of how such magnetically driven warps could lead to Lense-Thirring precession in NS-LMXBs.

\section{Summary}

The $\sim$1 Hz QPOs observed in dipping/eclipsing NS-LMXBs are caused by parts of the accretion flow that intercept our line of sight to the neutron star. The fact that these QPOs are only seen in dipping and/or eclipsing NS-LMXBs suggests that the material must have a high angular extent ($\sim$$20-30^\circ$) above the plane of the binary orbit. Such angles are close to the misalignment angles of 15--20$^\circ$ between inner disk and neutron star, required by the most recent Lense-Thirring precession models for LF-QPOs. The properties of the $\sim$1 Hz QPOs suggest that misaligned inner accretion flows are possible in NS-LMXBs, despite doubts that have been raised in the past.  It is unclear, however, if the $\sim$1 Hz QPOs and LF-QPOs are the result of the same precession mechanism. An inspection of all {\it RXTE} observations of the dipping/eclipsing NS-LMXBs will allow for a better comparison of the properties of the two types of QPOs.

\vspace{0.5cm}

I would like to thank the members of the Astronomical Institute `Anton Pannekoek' and SRON Utrecht, where part of this work was done, for their hospitality. I would also like to thank Maureen van den Berg, Diego Altamirano, Adam Ingram, and Chris Done for comments on an earlier version of this Letter. This work was supported in part by a NWO Visitors Grant. This research has made use of data obtained from the High Energy Astrophysics Science Archive Research Center (HEASARC), provided by NASA's Goddard Space Flight Center.


\end{document}